\documentclass[ams, 7pt]{revtex4}
\usepackage{amsmath, mathptmx} 
\usepackage{graphicx ,xcolor}
\usepackage{multirow,newlfont,enumerate} 
\usepackage{epsfig,epstopdf,ifpdf}
\usepackage{latexsym,slashed}
\usepackage{subfigure, float, epsfig}
\usepackage[final]{pdfpages} 
\usepackage{amsmath,amssymb,amsfonts,amsthm}
\usepackage{todonotes} 
\usepackage{tensor} 
\usepackage{titlesec} 
\usepackage{mathtools, array} 
\usepackage{todonotes} 
\usepackage{tensor} 
\usepackage{mathtools, array} 
\usepackage{mathtools, array} 
\theoremstyle{plain}%default 

\theoremstyle{definition} 

\newcommand{\beq}{\begin{equation}}
\newcommand{\eeq}{\end{equation}}
\newcommand{\beqr}{\begin{eqnarray}}
\newcommand{\eeqr}{\end{eqnarray}}

\def\wt{\widetilde}     
\def\lsim{\raise0.3ex\hbox{$\;<$\kern-0.75em\raise-1.1ex\hbox{$\sim\;$}}}
\def\gsim{\raise0.3ex\hbox{$\;>$\kern-0.75em\raise-1.1ex\hbox{$\sim\;$}}}
\def\para{\vspace{0.3cm}\noindent} 
\def\noi{\noindent}
\def\m{\,{\rm m}}
\def\s{\,{\rm s}}  
\def\g{\,{\rm g}}

\begin{document}

%%%%%%%%%%%%%%%%%%%%%%%%%%%%%%%%%%%%%%%%%%%%%%%%%%%
\def\wt{\widetilde}     
\def\lsim{\raise0.3ex\hbox{$\;<$\kern-0.75em\raise-1.1ex\hbox{$\sim\;$}}}
\def\gsim{\raise0.3ex\hbox{$\;>$\kern-0.75em\raise-1.1ex\hbox{$\sim\;$}}}
%%%%%%%%%%%%%%%%%%%%%%%%%%%%%%%%%%%%%%%%%%%%%%%%%%%%
\def\para{\vspace{0.3cm}\noindent} 
\def\noi{\noindent}
\def\m{\,{\rm m}}
\def\s{\,{\rm s}}  
\def\g{\,{\rm g}}

\title{Conformal folding Dynamics of Amyloid $\beta$ - $ \tau$ Oligomer  in Neuronal Dysfunction}

\author{ R. Dutta} 
\affiliation{Department of Physics, Material Science \& Astronomy, Missouri State University } 
\email{Corresponding Author :  rd5e@missouristate.edu / dutta.22@osu.edu} 

\author { A. Stan} 
\affiliation{Department of Mathematics, The Ohio State University  } 
\email{ stan.7@osu.edu}

\begin{abstract}
\end{abstract}
\maketitle 

\section{Introduction} 
 Alzheimer disease (AD) is neuro pathologically characterized by two hallmark lesions, which are extra cellular amyloid$\beta$ plaques and intracellular accumulation of abnormally phosphorylated $\tau$. 
Recent studies have shown that  the progressive accumulation of amyloid- $\beta$ plaque and neuro fibrility formation  in Alzheimer disease (AD) begins several  
years before the symptom onset ( i.e pre clinical stage) and follows distinct spatio - temporal pattern. They  exist in a great diversity of structure and morphology \cite{dear} and have been shown 
to play critical role in the formation of wide range of Amyloidogenic protein \cite{narayan} .Several  forms of amyloid protein  is known such as A$ \beta_{ 40, 42, 44} $ etc. 
  Amyloid precursor protein (APP) on neuron membrane shed constitutively A$\beta$ poly peptide as A$\beta_{40} $ and high levels of ROS( Reactive Oxygen Species) promotes  
  abnormal deposition of A $ \beta_{40} $ and longer  peptide  chain  A $ \beta_{42,44} $  \cite{seeman} which in turn produce another poly peptide chain called $\tau$. 
  Many clinical evidences [ \cite{scheff} and references therein] suggest that  there exists significant cross link  between A$\beta$ toxicity and $\tau$ pathology. 
   Gotz et al. \cite{gotz} reported fivefold increase  of neurofibrillary tangle (NFT) in cell bodies after injection of A$ \beta_{42} $ fibril into brain system of 6 month old  transgenic mice and 
    observed composition of NFT as twisted filament . Their data supported the hypothesis that A $\beta_{42} $ fibril can accelerate NFT formation  in vivo.
 Their observation found anatomical separation of amyloid induced axonal damage and possibly impaired axonal transport of $\tau$. 
  According to so called amyloid cascade hypothesis,  the progression of AD is related to presence of  soluble toxic oligomers of A$ \beta_{42,44} $ which are medium chain oligomer with molecular weight $ > M_{ \beta 40} $. 
 Amyloid $\beta$ and hyper phosphorylated $ \tau$ are considered as  hallmark lesions of Alzheimer disease. The loss of synapses and dysfunction of neuro transmission are directly tied to the disease severity.
 
 Amyloid oligomers are supramolecular structure consisting of several
Amyloidogenic protein non covalently assembled into matrix.   In general,  amyloid precursor protein (APP) ( monomer  of  $ \beta$  )  has been linked  to promote synaptic activity , synapse formation and dendritic spine formation which suggests that APP  plays pivotal role in learning and memory \cite{hoe}. 
  
The  hyper connectivity within DMN ( default mode network) is observed in cognitively unimpaired older adults with high levels of amyloid - $\beta$ but low in $ \tau$ or  high levels of both deposition.  
 IN healthy brain, under oxidative stress condition (  high rate of nucleation and association) , inflammatory signal ( e.g  high concentration of A$\beta$ ) gradually build plaque leading to neuron death $ n_d$ .

Most medium chain   bio polypeptide  are found in linear ( semi flexible) helical aggregate or coil form.  Dimer forms from lateral aggregation of two long chain whereas Tetramer ( helical aggregate of four  4 linear filament) 
 consist of two anti parallel dimer. Intermediate filament such as peptide oligomer have high tendency to dimerize or form tetramer in coiled fashion.  Oosawa [\cite{oosawa1} , \cite{oosawa2} ]
 discussed coil formation process as thermodynamic condensation process similar liquid - solid transition.  Oosawa et al and others   discussed  behavior of chain  aggregation in   conformal fold ,
   a natural consequence of thermodynamic condensation process depending on available free annealing thermodynamic energy in solution. 
    Herrmann et al \cite{herr} showed  that  unit - length filament (ULF) ( subunit of intermediate filament. of 32 -mer of length about 60 nm and diameter about 17nm ) in elongation phase 
    anneal longitudinally and fuse via their open ends .  Kirmse et al [\cite{kirm} and references therein]  demonstrated in  their mathematical model of cascade process that end - to -end anneal 
    is essential thermodynamic process for such polymer chain aggregation. 
Studies of various clinical data of amyloid and AD connection, it strongly suggests link between longer beta chain production and plaque formation. 
  Therefore, to study  connectivity between plaque matrix  at various neuronal site, one needs to develop model of neuron degeneration characterized by polymer conformal function within local volume. 
  In this paper, we develop model exhibiting relationship between conformal fold function and neuron degeneration rate within local brain volume in framework of semi flexible chain model and neuron connectivity. 

\section{ Mathematical/Physical Description  at Microscopic Scale } 

Various experimental data \cite{meyer1} , \cite{meyer2}  support plaque formation time few days in elderly patient. It suggests time scale of conformation can be fast or slow [ hours to years]. 
Main purpose of this paper is to provide a reliable mathematical model for macromolecule  with aggregate formation in solution  under favorable thermodynamic environment and related neuron death dynamics.  Main focus of this study is connectivity  between A$\beta$ and $\tau$  in terms of entangle  form. Both macromolecule have to be handled  via dynamics of end- to -end association or conformal fold form.  Finally, regulation of neuron dynamics should be affected by presence of such conformally folded protein.
We consider finite cerebral tissue region with smooth boundary ( $ V \in {\mathbb{R}}^3 $ , volume V  also represents  monomer solution local volume. This region contains several neurons with total neuron density $ n_0 $ . In the model, neurons are represented by smooth continuous dynamical density function ( a family of regular regions)  n (t,x)  and  progression of the disease described 
in terms of   equation of degenerated neuron density  $  n_d ( x, t, \tau) $ with $ \tau$ as time scale of connectivity between macromolecule dynamics and neuron degeneration  within $ V \times [ 0,T] $. 
  Degree of deterioration is associated with  degree of conformal folding state of both protein and cross link between them. We ignore diffusion of entangled polymer chain to other part of the brain as these are high mol weight filament in coil ( semi coil form) for which diffusive velocity is very low in typical brain fluid solution. rather, it affects only neurons in close region. 
So, plaque induced  neuron degeneration rate  equation  $ n_d (x, t, \tau) $ ( ignoring diffusion part)  can be   described : 
\begin{gather}
\cfrac{ \partial n_d  (x,  t, \tau)}{ \partial t}    =  J [ n_d ]   \\
J [ n_d ] =  \left[ \sum_{i}  d_{fni} \cfrac{{u_{i \alpha} (x, t, \tau )}^j }{ {u_{i   \alpha} (x, t, \tau ) }^j + k_i}  +  \sum_{j =0} d_{fnj} \cfrac{ {u_{j \alpha} (x, t, \tau )}^k }{ {u_{j \alpha} (x, t, \tau ) }^k + k_j } \right] n(t,x ) \\
\cfrac{ \partial n(t,x) }{ \partial t} =  \left[ a_0  - \cfrac{ a_1 n(t,x) } { s  + n(t,x)} \right]  * n(t,x ) - d n(t,x) 
 \end{gather} 
with vector valued function $ u_{i/j \alpha} (x,t, \tau ) $ representing  $\beta$ and $ \tau$ induced plaque function . Here    $  d_{fni/j}$ represent neuron death rate due to accumulation of polymer entangle of various length .
Parameters $a_0, a_1 $ and d represent production and death rate of neuron growth.   The delay time scale  
is defined  :   $ t = t - \tau $ with $\tau $ as connectivity scale between two macromolecule protein dynamics and disease progression. 
We ignore non local effect of degeneration as we assumed that within that local volume , local neurons will be affected due to presence of entangled polymer formation in the form of plaque. Since the process is totally thermodynamic process which depends on thermodynamic toxic condition and the gene to control monomer density within that region, non local effect will not contribute here.  Please note that following thermodynamics of polymer, it is the monomer density and free energy that gives rise to entangled polymer chain as thermodynamic process. 
The parameter j and k refer to degree of non linear accumulation of these two proteins in the sigmoid function. The death rate is function of correlation function between any two chain. 
Here,  we consider $ u_{i \alpha}  \equiv u_{i \alpha} ( t, x, \tau ) , i =1 \cdots N $  describing  to chain aggregate  with $ x \in V $ and N refers to 
maximal admissible length of polymer chain determined by chain density $c_0$ and Avogadro number $N_A$ as     $ N= c_0 V N_A$ . So, fluid  represents molar solution. 
In a a  closed system : 
\begin{gather} 
u_{ i/j \alpha}  ( x,t, \tau ) = {\sum_{i=0} }^\infty (ip)^\alpha  F_{ i/j \alpha  } (x, t, \tau ) \\
u_{i/j   f  \alpha }  (x, t,\tau) = {\sum_{i=0}}^\infty   i^\alpha  f_{i/j \alpha  } (x, t, \tau  )  
\end{gather} 
The parameter $ \alpha$ determines the degree of conformation such as dimer or tetramer ( two dimer in antiparallel) . Plaque formation / fibril production mechanism is described as 
time synchronization dynamics  between A$\beta$ oligomer and $ \tau$ macromolecule. The survival rate of neurons can be related to $ n(t) - n_d(t, x, \tau) $ at any time. 
The cognitive decline pattern will depend on degeneration rate and various other health factors. 
The evolution dynamics of $ F_{i /j} $ and /or $ f_{i/j} $ will be discussed in following  sections assuming evolution as polymer chain formation and entangle process. 
\vskip 6pt
\subsection{ Chain Growth and elongation as Filament} 
 Figure 2  illustrates  fibril formation and chain growth kinetics steps. 
 \begin{figure}
\includegraphics{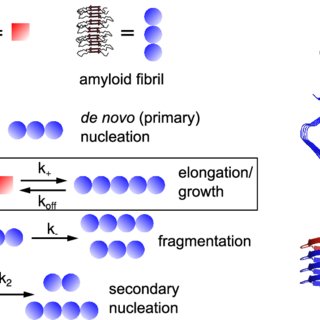}

\caption{Model for Fibril Formation as Polymerization Process }
 \end{figure}

 In relation to conformal folding and plaque formation, we start with   Smolchowski coagulation equation ( mean rate ) which describes  evolution of the mass/number spectrum of a collection of particles( here oligomer) due to successive merger.  If the masses of the particles are integral multiples of minimum mass  $m_1$ , coagulation 
 equation can be written in discrete form as: 
 \begin{gather} 
 \cfrac{d n_k}{ d t} = \cfrac{1}{2} \sum_{ i+j =k} A_{ij} n_i n_j - n_k \sum_{i=1} A_{ki} ni 
 \end{gather} 
 where $ n_k(t) $ is the number of particles of mass $ m_k = k m_1 $ in a volume V and $ A_{ij}$ is the rate coefficient ( or coagulation kernel) for merger between $m_i$ and $m_j$. In above equation , it is assumed that merging of $m_i$ and $m_j$ results in $m_i + m_j$ [ following mass conservation]. This equation can be written in continuous form 
 \begin{gather} 
 \cfrac{ d n(m)} { dt} = \cfrac{1}{2} {\int_0}^m d m^\prime A_{ m^\prime , m - m^\prime} n(m^\prime) n( m- m^\prime) \\
 - n(m) {\int_0}^{\infty} d m^\prime A_{m, m^\prime} n( m^\prime) 
 \end{gather} 
   Filament growth dynamics involve nucleation, elongation processes [\cite{abra}, \cite{oosawa1}, \cite{oosawa2}] . Within theoretical framework of equilibrium thermodynamics , chain formation , elongation and lateral association is multi step under favorable thermodynamic condition of monomer solution.  
  Formation of filament bundle through lateral association  can be  very quick  if  excess free energy of nucleotide bond is available in the solution.  In any assembly process, formation of intermolecular bond competes with greater translational and rotational entropy of monomer in solution. 
 
 It is necessary that   monomers  come close together using  available free energy to form primary seed nucleus  ( determined by number $n_c$ , defined as critical size) 
  of  polymer chain and 
 any two bead( seed )  nucleus can join linearly  to form  polymer chain ( rigid or flexible ) in next time. 
There exists possibility of secondary nucleation at various surface site. 
 Masel et al \cite{masel} have shown  that most peptide bond polymer such as 
 PrP polymers are formed by one dimensional 
 aggregation of primary nucleus unit .  Therefore, we ignore  secondary nucleation pathways.  A filament of length i, $ f_i$ is formed due to addition of m(t) to $ f_{i-1} $ or by end- to end 
association of two shorter filaments ( $ f_j $ and $ f_k$ , where i = j + k ).   Elongation process can be described as addition of monomer to both open end of  a chain. 

 In our model ,  filament initiation assumes reversible self adhesion of monomer function m(t) to form oligomer nucleus ( $ N (t) \propto n_c    m_t - m(t)  $, where $ n_c $   refers concentration of monomer  in primary seed nucleus.      In a solution of initial monomer concentration  $ m_t , M(t) = m_t - m(t) $ is the monomer concentration used in polymerization process. 
  Within framework of Oosawa model of kinetic theory of polymerization [ \cite{oosawa1}, \cite{oosawa2} ] ,  evolution dynamics of single chain function , denoted by $ f_i(t) $  can be described as multi step cascade process :
 \begin{gather} 
 \cfrac{ \partial f_i(t)}{ \partial  t} =  k_n m(t)^{n_c}   \delta_{ i, n_c}   +  2 k_{+} m(t) \left[ f_i(t) - f_{i -1} (t) \right] + 2 k_{off} \left[ f_{i + 1}  (t) - f_i (t) \right]  \\
 - p k_l f_i (t) \sum_{j=1}  {f_j(t)}^{p-1} 
 - \sum_{j=n_c+1 }  {k_{i, j}}^\prime  f_i(t) f_j(t)  +  \cfrac{1}{2} \sum_{ j=n_c +1 } {k_{i, j } }^\prime f_i (t) f _j (t)    \\
 f(t) = \sum_i f_i(t) \\
 \cfrac{\partial N(t)} { \partial t} =  - k_n m(t)^{n_c}   \delta_{ i, n_c}  - 2 k_{+} m(t) \left[ f_i(t) - f_{i -1} (t) \right]  
   \end{gather} 
followed by mass/number preservation principle 
  \begin{gather} 
   \cfrac{\partial m(t)}{ \partial t} = - \cfrac{ \partial }{ \partial t} \sum_{ i=n_c} i f_i (t) 
\end{gather} 
Here $ t \equiv t - \tau $ . 
 
The parameters    $  k_n, k_{+} $ and $k_{off} $ refer to  equilibrium nucleation, elongation and  dissociation rate respectively.  In this work, we develop model for lateral assembly of single chain through longitudinal bond to form ULF ( with degree of lateral association as p , same as number of chain ).  The maximal length of a stretched ULF as N within given volume V will be  determined by $ N = c_0 V N_A$ , with  $ c_0 $  is the initial concentration of such ULF unit, $N_A $ as Avogadro number in solution. In bulk phase, such ULF unit can associate end-to end  to produce various conformal structure  depending on contour length and radius of gyration.  The aggregate structure will depend on  
end-to -end rate constant and correlation function.

 \section{Calculation of  Aggregate Rate Constant  }  
 \subsection{ Lateral Association Rate Constant} 
  Based on brownian dynamics diffusion  of motion under influence of electrostatic interaction ,  several authors developed model for   protein oligomerization, enzyme analysis and other similar kinetics   to obtain analytical expression for protein - protein lateral association rate constant including significant work by  Solc. \& Stockmayer \cite{solc}. 
  
In the model,  each protein is  assumed to be spherical that bear point charge and undergo translational as well rotational Brownian motion  under mean electrostatic potential.  Two spheres should be properly aligned ( via rotational diffusion)  before they associate with each other  needed to  form complex. 
   Based on diffusion model given by  Smolchowsky  for two uniformly reacting  sphere with centrosymmetric interaction potential U(r) ,    association rate follows  : 
   \begin{gather} 
    k_{laij} =  4 \pi D {\int_R}^{\infty} e^{ \cfrac{ U(r) }{ k_B T} } r^{-2} d r
   \end{gather} 
   where D is the  steady state  diffusion constant ( basal) ,  determined by   $ D= D_i + D_j $; sum of individual translational diffusion constant. This model lacks essential orientational alignment 
 of both spheres. 
 To include  orientational adjustment to form complex in protein, 
    Solc \& Stockmayer \cite{solc}  studied using Quasi - Chemical analysis and gave  analytical  form for $k_{laij} $ ( any two chain i and j coming and bonding laterally 
   within angle angle  between 0 and $ \delta_i $  ) as : 
    \begin{gather}
  k_{laij} =4 \pi D R_{ij}  \left[  F_i \xi_j \ tan ( 0.5* \delta_j) + F_j \xi_i \ tan ( 0.5* \delta_i) \right] 
 \end{gather}  
  The function $ F_i  ( i=1 or 2 ) $ represent surface fraction of individual sphere within reactive patch i.e surface between  0 and $ \delta_i $ .
 Here,  $ \xi_i =  \sqrt{ \cfrac{( 1 + \cfrac{D_i R^2  }{ D} } {2} } $ with   $ D_i $ as individual  orientational diffusion constant.  An approximate estimation , as obtained by Berg \cite{berg} for $ F_i $ in case of patch within angles between 0 and $ \delta_i $ is given by : 
 \begin{gather} 
 F_i = \cfrac{ \xi_i + \ cot ( 0.5 * \delta_i) }{ \xi_i + \ sin( 0.5* \delta_i) \ cos ( 0.5 * \delta_i) } 
 \end{gather}

\subsection{End - to - end association Rate Constant}

Based on brownian dynamics of collision of any two approaching spheres, Hill \cite{hill} gave  length dependent end- to end association rate constant  relation  with  probability distribution for mutual orientation  $ p_{ij} $ as: 
 \begin{gather} 
k_{ij} = 4 \pi ( D_i + D_j) R_{ij} p_{ij} \\
D_i = \cfrac{ k_B T}{ 3 \pi \eta L_i} \left( 1 + \sqrt{ \cfrac{ 6}{ \pi L_i} } p(Li) \right) 
 \end{gather}
 Here ,  $\eta$ refers to  viscosity of the solution.  Hill \cite{hill} showed that probability  $ p_{ij} $ of such  contact within interaction zone  is inversely proportional to interaction radius i.e 
\begin{gather} 
p_{ij} \propto \cfrac{1}{ {R_{ij}}^2} = \cfrac{p}{ {R_{ij}}^2}
\end{gather} 
which indicates that two sphere have to be in contact with each other. 
 For end-to -end annealing , $R_{ij} $ depends on  sum of end-to end mean square distance of individual chain . Following WLC theory  of semi flexible chain of a macromolecule , mean end - to -end distance of a flexible chain follows 
\begin{gather}
\langle {R(L_i )}^2 \rangle = l_p L_i - 2 {l_p}^2 \left( 1 - exp( - \cfrac{L_i}{2 l_p} ) \right) 
\end{gather}
with  $l_p$ is defined as  persistence length of a chain. 
Persistence length characterizes  flexibility  or orientation of a chain : 
 \begin{gather} 
 l_p = - \cfrac{1}{ \ ln ( \ cos \delta_{i/j}  ) } 
 \end{gather}
with $\delta_i $ is the orientation  angle of i th chain with respect to origin. 
 $ R_{ij}$ can be approximated by half of total end - to -end distance of chain i and j , given by 
\begin{gather} 
R_{ij} \approx \cfrac{\langle {R(L_i )}^2 \rangle + \langle {R(L_j)}^2 \rangle}{2} = \sqrt{  l_p L_i - 2 {l_p}^2 \left( 1 - exp( - \cfrac{L_i}{2 l_p} ) \right) } + \sqrt{ l_p L_j - 2 {l_p}^2 \left( 1 - exp( - \cfrac{L_j}{2 l_p} ) \right) }
\end{gather} 
Based on maximum entropy principle on continuous chain model, Harnau et al \cite{harnau} studied Worm like chain similar to Kratky - Porod chain and studied structure function of such chain movement in solution with arbitrary stiffness. They estimated diffusion co efficient of  a  macromolecule from time correlation function in translational mode as 
\begin{gather} 
D_i = \cfrac{k_B T} { 3 \pi \eta L_i}  \Biggl \{  1 + \sqrt{\cfrac{ 6}{ \pi L_i } } {\int_{d_i}}^{L_i }  d s \cfrac{ L_i - s } { \sqrt{a(s)} } exp \left[ - \cfrac{ 3 {d_i}^2} { 2 a(s) } \right]   \Biggr \}
\end{gather} 
where  $d_i$ refers to stokes diameter of i th conforming chain.
\begin{gather} 
a( s - s^\prime) = \cfrac{ \mid s - s^\prime \mid } { 2 l_p} - \cfrac{1}{ 2 {l_p}^2} \left( 1 - exp( - \cfrac{  \mid s - s^\prime \mid }{ l_p} \right) 
\end{gather}

 Replacing expression  of a(s) into equation (11) and integrating equation  we obtain  solution for end - to -end conformal rate constant as :

\begin{gather} 
k_{ ij} = p \cfrac{ 4 k_B T}{ 3 \eta} \cfrac{ N_A}{ 1000} \left[  \cfrac{1}{L_i} \left(  1 + \sqrt{ \cfrac{6} { \pi L_i} }  a(L_i )  \right)  + \cfrac{1}{L_j} \left(  1 + \sqrt{ \cfrac{6} { \pi L_j} }  a(L_j )  \right) \right] 
\end{gather} 
The function $ a(L_{ i/j} ) $ can be obtained integrating equation (12) as: 
\begin{gather}
a(L_i) = {\int_{d_i}}^{L_i }  d s \cfrac{( L_i - s) }{ \sqrt{ l_p L_i - 2 {l_p}^2 ( 1 - e^{ - \cfrac{L_i}{ 2 l_p}  } ) } + \sqrt{ l_p L_j- 2 {l_p}^2 ( 1 - e^{ - \cfrac{L_j}{ 2 l_p}  } ) } } 
\end{gather} 
Values of lateral and end- to - end association rate constant have to be implemented into chain aggregation dynamics .

\section{ Conformal Folding Dynamics} 
 Most experimental data ( light scattering image and other imaging technique) confirm of solid plaque as non soluble structure  at various neuron site in Alzheimer patient . Therefore, to understand plaque formation as polymer conformal structure , we develop  model of  conformal folding  aggregation of chain dynamics  in two steps. Following are the steps: 
  First step: when a chain or filament is created of length i as $ f_i(t) $ into solution following kinetics of Oligomerization of monomer under available  thermodynamic free energy increment i.e ,  $ s= exp( -  \partial f^{\prime} / k_B T ) $  necessary for 
  such  i - mer entangle  formation  from   p i  -mer filament lateral  association. In steady state condition, two or more single chain  come closer under attractive electrostatic potential . As a result , they aggregate via lateral association to form bundle or ULF ( a unit lateral filament). Second step , these ULF  anneal together  via end - to - end association. These long oligomer  bundle (ULF)  then  conform to a coil or coil like structure if thermodynamically favorable under chemical condition of solution within brain volume ( discrete) . These coil like folded structure is generally   insoluble irreversible structure under steady state thermodynamics. 
 To study assembly and folding dynamics of elongated objects,   It is essential  to study their length distribution as length distribution influences lateral association rate which in turn influences conformal folding rate constant.  Several short filaments 
 will contribute much more to polymerization than one long filament. Therefore, to study oligomer ( intermediate filament ) assembly , temporal evolution of $ F_i (t) $ , concentration of entangle variable of degree i should be described . 
 It is assumed formation an association rate constants are independent of length . So, formation of j filament will be : 
 Following coagulation theory of Smolchowski , merging dynamics of ULF  , represented by function $F_i(t)$  in continuum system   can be written  : 
 \begin{gather} 
 \cfrac{ \partial F_i}{ \partial t} = \cfrac{1}{2} {\sum_{ j=1}}^{ i-1} k_{j , i -j } F_j F_{ i -j} - F_j {\sum_{j=1}}^{ N - i} k_{ i j} F_j  \\
 i = 1 \cdots N 
 \end{gather} 
 with  N as maximal possible length within given volume  i.e  $ N = c_0 V N_A $, $ N_A $ being  Avogadro number.  We assume here that lateral aggregation is thermodynamically favorable, instantaneous and irreversible mechanism compared to fragmentation and dissociation. 
 Various works such as  [ \cite{ fumio} and the references therein] have supported helical aggregation of macromolecules in equilibrium condition like solid - liquid transition phenomena instead of linear aggregation. 
In linear polypeptide, \cite{zimm} have that shown transition phenomenon from intramolecular helix to random coil is similar to  solid- liquid transition or like melting phenomenon. Unlike oosawa work \cite{oosawa1},  consider   that when p number of such oligomer of length i  come in  contact with each other within given volume, they form  nonhelical 
bundle of  unit of $F_i(t) $ (  ULF , unit lateral  function) .
 Considering q as degree of such lateral adhesion polymerization ( also called  longitudinal degree of polymerization) , evolution dynamics of $F_i(t)$ follows relation: 

\begin{gather} 
\cfrac{ \partial F_i (t)}  { \partial t} =  - k_{l a} p f_i (t) \sum_{j=1}  {f_j(t)}^{p-1}   +  \cfrac{1}{2} {\sum_{ j=1}}^{ i-1} k_{j , i -j } F_j F_{ i -j}  - \sum_{j=i}  k_{ij} F_i(t) F_j (t)  
\end{gather} 
Here ,  function $ F_i(t) $ represents entangle of various order. The last two terms on RHS represent conformally folded entangle by end-to end association. 
 One can calculate mean length of the bundle length distribution for N such $ F_i(t) $  for end- to -end adhesion dynamics : 
 \begin{gather} 
 \langle L_N(t) \rangle := \cfrac{ {\sum_{i=1}}^N i F_i(t) }{ {\sum_{i=1}}^N F_i (t) }
 \end{gather} 
 which is interpreted as mean degree of conformal polymerization .

  as moment function of   nucleation , elongation and aggregation evolution dynamics .  Evolution  dynamics   can be described mathematically as: 
 
 \begin{gather} 
\cfrac{u_{i/j f \alpha } (x, t,\tau)  }{\partial t} =  - k_n m(t)^{n_c}  {\sum_{i=n_c+1}}^{\infty} i^\alpha +  k_{+} m(t) {\sum_{i=n_c+1}}^{\infty} i^\alpha \left[ f_i (t) -
 f_{i-1} (t) \right]  \\
  - p k_{laij}  {\sum_{i=n_c+1}}^{\infty} i^\alpha   f_i(t) \sum_{j= 1} {f_j (t) }^{q-1}  - \left[  \cfrac{1}{2}  {\sum_{i=1}}^{\infty} i^\alpha  \sum_{ j= 1} \sum_{k= i - j} k_{ j, i-j} i^\alpha  f_j(t) f_{i-j} (t)  
  -  \sum_{j= n_c+1} k_{ i j} i^\alpha  f_i (t) f_j(t) \right]   \\
   \cfrac{\partial m(t)}{ \partial t} = - \cfrac{ \partial }{ \partial t} \sum_{ i=n_c+1 } i f_i (t)  \\
   \cfrac{ \partial u_{i/j    }  (x, t,\tau)} { \partial t} = {\sum_{i=n_c+1}}^{\infty} i^\alpha  \left[  \cfrac{1}{2} {\sum_{ j=1}}^{ i-1} k_{j , i -j } F_j F_{ i -j}  - \sum_{j=i}  k_{ij} F_i(t) F_j (t)  \right] 
\end{gather} 
with $ n_c $ as oligomer nucleus size.  Following oligomer formation through primary nucleation process , $n_c$ is the number of monomer per nucleus. When these nucleus adhere linearly, a filament is produced. Filament can be of any size i from $ ( n_c + 1) $ as minimum length to $ \infty $. Due to monomer addition at both end of filament , there exist 2i monomer per filament $f_i(t) $ function.  Considering helical and conformal aggregation process as natural kinetic phenomenon during filament growth process , decrease in filament growth should account for increase in helical and/or 
coil growth in kinetics . 
Here, we  assume that the solution of macromolecule  have the ability to make linear ( coil form ) aggregate( linear polymer chain ) by end - to - end association.  The solution contains 
 dispersed macromolecules ( monomer) and linear polymer of various length ( various degree of polymerization) following mass - action principle.    
 similar to gas- liquid condensation process in which dispersed monomers correspond to gas and helical molecule as liquid molecule as natural process of polypeptide protein molecule.     
 We assume  that  f- type oligomer [ denoted by $ f_i $ of any length can laterally adhere with other of same or larger  length and lateral assemble of p oligomers will have heterogeneous mass unit length along respective filament. Thus, $ u_{f 2 } (t) $ function or  one dimer ULF function refers to concentration of   dimer  $ f_1(t) f_1(t) $  with $k_{la 11} $ as dimer formation rate constant in equilibrium to form end-to -end conformal function 
 through  end- to -end association  of two smaller oligomer $ f_{i -j }$ with $ f_j $ will produce an linear conforming chain of length i .  The equilibrium rate constant for this process ,  $ k_{ij} = k_{1 1} $ 
 refers to  adhesion of two ULF unit  $ F_1(t) F_1(t) $  represented by $ u_{F 2} (t) $. 
 With this idea of aggregation to form dimer, tetramer etc, set  $ \alpha=2$ . Then chain dynamics equations can be expressed : 
\begin{gather} 
 \cfrac{ \partial u_{ f 0 } (t) } { \partial  t} =  k_{+} m(t)  N(t) -  p k_{la ij }  {u_{ f 0} }^q - \cfrac{1}{2} k_{ij}  { u_{ f 0} }^2 \\
 \cfrac{ \partial  u_{ f 1} (t) } { \partial  t}  =  k_{+} m(t) \lbrace ( n_c+1) N(t) + u_{ f 0 } \rbrace - p k_{la}  u_{ f 1 } {\lambda_{ f 0} }^{q -1}  \\
\cfrac{ \partial  u_{ f 2} (t) } { \partial  t}  =  k_{+} m(t) \lbrace ( n_c +1)^2 N(t) + 2 \lambda_{ f 1} + u_{ f 0 } \rbrace 
\end{gather} 
 with 
 \begin{gather} 
 \cfrac{\partial m(t) }{ \partial t} = - k_n n_c  {m(t)}^{n_c} -  k_{+} m(t) \left[ f(t) +  u_{ f 0 } (t) \right] \\
 \cfrac{ \partial N(t)}{ \partial t}  = k_n   {m(t)}^{n_c} - k_{+} m(t) N(t) 
 \end{gather} 
where N(t) is defined evolution of  oligomer nucleus concentration at any time t.  One can interpret  equation (34 - 35 )  as nucleus and single chain density  function in solution respectively as defined in equation (4). 
 
 Free monomer density function m(t) can always  be related to  $ M(t) = m_{tot} - m(t)$ as monomer density used in polymerization process. 
 In any dilute solution of free monomer , monomer density of any oligomer obeys hyperbolic functional relationship with nucleation , dissociation and elongation rate as : 
  \begin{gather} 
m(t) =Rootof \left\{ t \sqrt{ 4 k_{+} k_n {c_0}^{n_c} n_c } -  2 arctanh\left(  \cfrac{\sqrt{  - k_{+}{ k_n}^{n_c} n_c +  2 k_{+} k_n {c_0}^{n_c} n_c }}{ 2 \sqrt{ k_{+} k_n {c_0}^{n_c} n_c } } 
 \right) \right\} 
\end{gather}

The equation for NFT formation dynamics , represented by $ u_{ F 2} (t), u_{ F 1 } (t)$ and $ u_{ F 0 } (t) $ of various order ( determined by $\alpha$ ) gives information about protein folding  dynamics . Following equations can be solved to obtain solution for NFT tangle up to second order  
 \begin{gather} 
 \cfrac{ \partial  u_{ F 0 } (t) } { d t} = k_{laij } {u_{ f 0} }^q - \cfrac{1}{2} k_{ij} \ {u_{ F0} }^2  \\
  \cfrac{ \partial u_{ F 1 } (t) } { d t} = p  k_{la} u_{ f 1} (t) {u_{ f 0} }^{q -1} \\
    \cfrac{ \partial  u_{ F 2} (t) } { d t} = p^2  k_{la} u_{ f 2} (t) {u_{ f 0} }^{q -1} + k_{ij}  {u_{ F 1 }}^2 
 \end{gather}

Conformal folding  rate constant  can also be   determined  based  on internal end - to -end distribution function $  f( \xi) $ , a central quantity to characterize conformation of single polymer chain 
\begin{gather} 
f( \xi) = 2 \pi \xi G( \xi) 
\end{gather}

$ \xi(L) $ is the  contour length of a given chain of length L. It is the minimum distance between s = 0 and s =L  for a given chain length. 
 
 In flexible/semi flexible limit,  $ G( \xi) $ can be calculated numerically   \cite{jan} as:
 \begin{gather} 
 G( \xi) = \cfrac{1}{ 4 \pi \mathbb{N} } \cfrac{ \kappa} { 2 \sqrt{\pi}} \sum_{l=0} \cfrac{1} { {\kappa( 1- \xi)}^{3/2} } \times  exp \left[ - \cfrac{{ ( l -0.5)}^2}{ \kappa(1 - \xi) } \right]  H_2 \left[ \cfrac{ l -0.5} { \sqrt{ \kappa( 1 - \xi) } } \right] 
 \end{gather} 
where  $ H_2(x) = 4 x^2 - 2$ is the second order Hermite polynomial. The above series is shown to converge for l= 1 for $ \kappa ( 1 - \xi) \le 0.2 $ [ flexible limit]
 $ ; \kappa$ being bending modulus. 
 For model like WLC ( Worm like Chain model ) which considers only short range interactions between neighboring monomer , $ G(\xi) $ is the probability density of finding any two monomers at average relative position $ \xi = \xi(s) - \xi( s^\prime) $. 
Within short range interaction like WLC model,  mean end - to - end vector length as a function of each contour length s is given by 
\begin{gather} 
\langle \xi^2 (s) \rangle  = 4 l_p s \left[ 1 - \cfrac{ 2 l_p}{s} \left( 1 - exp\left( - \cfrac{s}{2 l_p} \right) \right) \right]  \
\end{gather}
with $l_p$ , known as Kuhn length  characterizing flexibility or bond orientation between neighboring monomer.
For flexible and semi flexible range, WLC model does not exhibit proper behavior of above relation and scaling behavior is found to be more suitable for stiff to flexible polymer following \cite{reche} as: 
\begin{gather} 
\langle \xi^2 (s) \rangle \propto \cfrac{ s^{2 \nu} {( L - s)}^{2 \nu} } { s^{2 \nu} + { ( L -s) }^{ 2 \nu} } 
\end{gather}
for a given chain length L . The value of the parameter $ \nu$ varies in different regime of s , such as linear( stiff rod ) , semi flexible and flexible zone.
    Since nature provides different length scale of different stiffness such as  $ l_p \approx 10 $ nm for spectrin, $ l_p \approx 50 $ nm for DNA , $ l_p \approx 17 \mu $m for actin and $ l_p \approx 5.2 $ nm for microtubule, we will use length scale accordingly.

\subsection{ Chain Growth Dynamics of   amyloid $ \beta $ and $ \tau$ Oligomer  } 
Several experimental data suggest that both amyloid $\beta $ and $ \tau$ protein are linear oligomer macromolecule. This indicates nucleation takes place only via homogeneous primary pathways. 

We  consider   $  \beta_m(t) $ and $\beta_o(t) $ to be  concentration of $\beta$  monomer and seed nucleus  within brain  volume. 
 Since biopolymer entangle depends on elasticity ( bending ) of the chin with two significant length scale , contour length l and persistence length $ l_p$ . 
 When confined to surface or volume of  a sphere , radius of confinement r plays significant to determine conformal structure. 
 r ( contour length)  is determined as end-to-end distance of a polymer chain depending on bond strength and angle of confinement. The equation set for amyloid - $\beta$ chain  dynamics will be described:
 \begin{gather} 
 \cfrac{ \partial  u_{\beta f  0 } (t) } { \partial  t} =  k_{+} \beta_m (t)  \beta_o(t) -  p k_{la ij }  {u_{ \beta f 0} }^q - \cfrac{1}{2} k_{ij}  {u_{ \beta f 0} }^2 \\
 \cfrac{ \partial  u_{ \beta 1} (t) } { \partial  t}  =  k_{+} \beta_m (t) \lbrace ( n_1 +1) \beta_o(t) + u_{ \beta f 0 } \rbrace \\
  - p k_{laij }  u_{ \beta  1 } {\lambda_{ \beta f 0} }^{q -1}  \\
\cfrac{ \partial  u_{ \beta f 2} (t) } { \partial  t}  =  k_{+} \beta_m(t) \lbrace ( n_1 +1)^2 \beta_o(t) + 2 u_{ \beta f 1} + u_{ \beta f 0 } \rbrace \\
\cfrac{\partial \beta_ m(t) }{ \partial t} = - k_n n_1 {\beta_m(t)}^{n_1} -  k_{+} \beta_m(t) \left[ \beta_o(t) +  u_{ \beta f 0 } (t) \right] \\
\cfrac{ \partial \beta_o(t) }{ \partial t} = k_n   {\beta_m(t)}^{n_1} - k_{+} \beta_m(t) \beta_o(t) 
 \end{gather} 
with $ n_1$ representing amyloid oligomer nucleus size. 
 Here total concentration of monomer to trigger amyloid $\beta $ oligomerization process is $ \beta_{tot}  \approx \beta_m(0) $  

It is well studied that certain oxidative stress local in brain triggers excessive production of amyloid $\beta$ oligomer.  The excessive amyloid protein beyond normal concentration, in turn 
triggers production of  another oligomer , called $\tau$ macromolecule. Both macromolecule in conformal form are held responsible for neuronal degeneration and associated cognitive decline with time. The process can be modeled as time delay mechanism which we can associate as early onset of degeneration or late onset . 
 We assume $\tau_m(t) $ and $\tau_o(t) $ as tau monomer and oligomer nucleus concentration available in the solution and growth dynamics can be obtained through 

\begin{gather}
 \cfrac{ \partial \tau_o(t)}{ \partial t}  = k_n   {( \beta_m - \beta_{th}) (t - \tau) }^{n_2} - k_{+} \beta_m(t) \tau_o(t) \\
 \cfrac{\partial \tau_ m(t) }{ \partial t} = - k_n n_2  {( \beta_m - \beta_{th}) (t - \tau) }^{n_2}  -  k_{+} \tau_m(t) \left[ \tau_o(t) +  u_{ \tau f 0 } (t) \right] \\
\cfrac{ \partial  u_{\tau  f 0 } (t) } { \partial  t} =  k_{+} \tau_m (t)  \tau_o(t) -  p k_{la ij }  {u_{ \tau f 0} }^q - \cfrac{1}{2} k_{ij}  {u_{ \tau f 0} }^2 \\
 \cfrac{ \partial  u_{ \tau f 1} (t) } { \partial  t}  =  k_{+} \tau_m (t) \lbrace ( n_1 +1) \tau_o(t) + u_{ \tau f 0 } \rbrace \\
  - p k_{laij }  u_{ \tau f 1 } {u_{ \tau  f 0} }^{q -1}  \\
\cfrac{ \partial  u_{ \tau f 2} (t) } { \partial  t}  =  k_{+} \tau_m(t) \lbrace ( n_1 +1)^2 \tau_o(t) + 2 u_{ \tau f 1} + u_{ \tau f 0 } \rbrace \\
 \end{gather} 
with $ n_2$ representing $\tau$  oligomer critical nucleus size.

 \subsection{ Entangle Growth Dynamics} 
 
 NFT dynamics of both protein macromolecule  can be obtained solving  following equations 
 \begin{gather} 
   \cfrac{ \partial  u_{ \beta F 0 } (t) } { \partial t} = k_{laij} {u_{ \beta f 0} }^q - \cfrac{1}{2} k_{ij} \ {u_{ \beta F0} }^2  \\
  \cfrac{ \partial  u_{ \beta  F 1 } (t) } { \partial  t} = p  k_{laij} u_{ \beta f 1} (t) {u_{ \beta f 0} }^{q -1} \\
    \cfrac{ \partial u_{ \beta F 2} (t) } { \partial  t} = p^2  k_{laij } u_{ \beta  f 2} (t) {u_{\beta  f 0} }^{q -1} + k_{ij}  {u_{ \beta F 1 }}^2  \\
    \cfrac{ \partial  u_{ \tau F 0 } (t) } { \partial t} = k_{laij} {u_{ \tau f 0} }^q - \cfrac{1}{2} k_{ij} \ {u_{ \tau F0} }^2  \\
  \cfrac{ \partial  u_{ \tau  F 1 } (t) } { \partial  t} = p  k_{laij} u_{ \tau f 1} (t) {u_{ \tau f 0} }^{q -1} \\
    \cfrac{ \partial u_{ \tau F 2} (t) } { \partial  t} = p^2  k_{laij } u_{ \tau f 2} (t) {u_{\tau f 0} }^{q -1} + k_{ij}  {u_{ \tau F 1 }}^2 
\end{gather}    
Here chain dynamics of both oligomer need to be solved in coupled manner and  NFT dynamics follow individual intramolecular of heterogeneous chain or similar macromolecule  entangle growth mechanism.

\subsection{Calculation of Conformal Structure Function in multi chain aggregate} 
Conformal structure function  ( shape of the chain) for  polymer fibril ( multi chain aggregate) can be related to mean square end-to -end distance function. It takes care of the structure related to $ cos( \theta ) $ , where $ \theta $ is angle between two segment or filament. 
 b. Then, 

It can be shown that probability for end - to - end conformation structure given by $ R_N $ of 2D ideal chain is 
\begin{gather} 
P(N, \vec{R}_N)   = \cfrac{3}{ 2 \pi N b^2} exp\left( - \cfrac{ 3 {R_N}^2 }{ 2 N } \right) 
\end{gather} 
The WLC model is considered  as most studied flexible polymer chain model and can be viewed as a limiting case of free rotating chain . 
The model yields end-to end mean square distance  of the chain 
\begin{gather} 
\end{gather}
In the limit $ l_p <<  L $ , one gets $ \langle {R_N}^2 \rangle = 2 l_p L $ and the chain behaves like Gaussian coil ( through end- to - end association) . For $ l_l > > L$, 
$ \langle {R_N}^2 \rangle = L^2 $ and chain behaves like rod. For fibril through lateral association like alpha helix without cross over ,  Most bio polymer such as various protein , DNA behave as semi flexible polymer. Semi flexible polymer behave like rod on small scale. For larger length scale, entropic flexibility occurs and random coil like structure occur. WLC model envisions a continuously flexible isotropic rod ( freely jointed) 
scale and form twist structure easily within finite volume. The twist structure depends on $ \cfrac{l}{b} $ ratio typically , l being single chain length and b  being monomer size. According to In dilute to semi dilute regime, varieties of filament structure forms . In framework of Monte Carlo simulation studies [ based on bond fluctuation theory], there exists relationship between bond length $ b_e $ and root mean square end - to - end distance

The production of entangled polymer chain and plaque in neuron system that controls  neuro degeneration and cognitive decline process  will be governed by each chain dynamics equation set. 
The appearance of correlation coefficient  in the neuro degenerative set of equations  ( $ \beta - \tau$ and $ \tau \tau $ dimer  )  have to be solved using chain dynamics in continuum limit. 
The main idea is to use a pseudo continuous model of a polymer solution ( synaptic fluid in brain ) , consisting of long chain. 
Each chain is normally endowed with N monomers linked by linear spring interaction of length b. The energy configuration of the cth chain represented by 
a continuous path $ r_c( t,s) $ with monomer label  $ s \in (0,1) $. The chain dynamics and correlation coefficient foe chain entanglement can be expressed in terms of end - to end distance vector between monomer as

\begin{gather} 
\cfrac{ \langle R^2(n) \rangle} { n l_o} = l_k \{ 1 - \cfrac{ l_k}{2 n l_o} \left[ 1 - exp \left( - \cfrac{ 2 n l_o}{l_k} \right) \right] \}
\end{gather} 
where $ \langle R^2 (n) \rangle $ is the mean - squared distance between monomers separated by peptide bond distance n and $ l_o  $ is the backbone length( persistence length ) . 
Packing fraction is defined in terms of number density of chain \cite{fetters} as $ p = \cfrac{1}{ \rho_c \langle {R_{ee}}^2 \rangle } $ 
Here $ \langle  {R_{ee}}^2 \rangle $ denotes end-end distance of the chain which can be provided  as input.  We implement idea of volume fraction within given volume [ tube theory] that each chain occupies \cite{hoy}  as 

\begin{gather} 
\Omega = N \cfrac{ \phi}{\rho} = \cfrac{\phi}{ \rho_c} = \pi { ( \cfrac{ d}{2})}^2 N l_o 
\end{gather} 
where d defines effective chain diameter which is different from tube diameter a. The maximum diameter or tube diameter is defined as end-to-end distance of a Gaussian chain of chemical length $ N_e$ and effective bond length b as
\begin{gather} 
b= \sqrt{ l_k l_o}  \\
N_e =  \cfrac{a^2 }{b^2 } \\
L_e = N_e  l_o
\end{gather}
and $ \theta $ can not take $ 0^\circ$ in order to have twisted filament. 
The role of bond angle will determine various conformal structure of twisted filament and plaque matrix.  

\section{Numerical Simulation and Results}  
  Numerical solution of above coupled equations   exhibit exponential growth  with saturation value required to trigger the onset of the disease [ Figure I]. 
 Figure I exhibits  fibril growth with no time lag between $ \tau_o$ trigger and fibril formation and shown below
\begin{figure}
\includegraphics[scale=0.35]{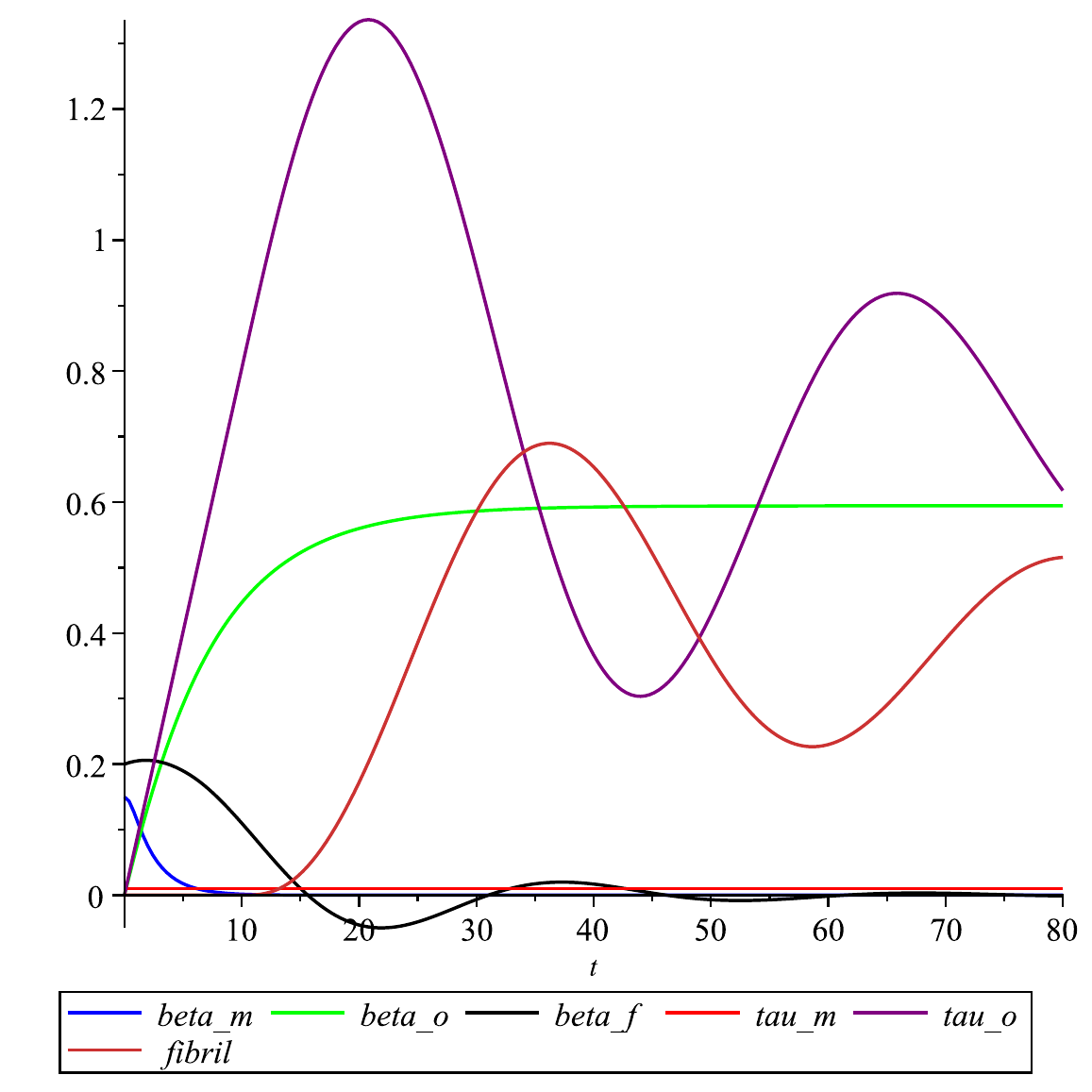}
\caption{Biomarker growth with no time lag  }
 \end{figure}
 
Figure 2  exhibits biomarker growth of two oligomer and fibril and entangle formation  with time lag. 
\begin{figure}
\includegraphics[scale=0.35]{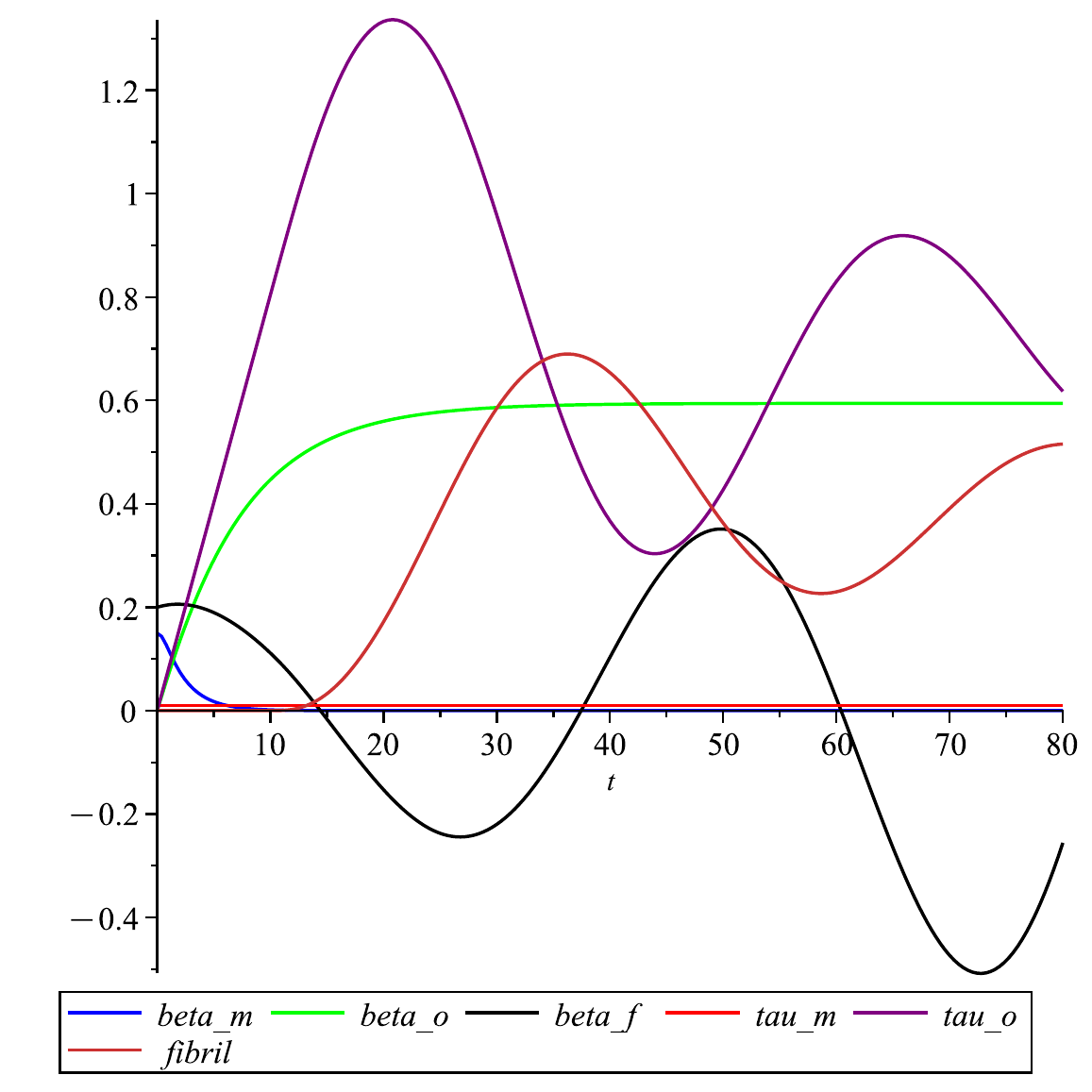}
\caption{Biomarker growth of $\beta$ and  $ \tau$ formation  and entangle with time lag apprximately 5.04  }
 \end{figure}
. 
 
\ The dynamics for neuronal dysfunction neuro(t)  and corresponding cognitive decline function cog(t)   can be represented  as coupled dynamics 
 \ In both cases , we observe very high growth rate of phosphorylated $ \tau$ production which is considered as hallmark of AD progression. 
 Simulation results for neuro degeneration function is shown in following plot [ Figure 3] 
\begin{figure}
\includegraphics[scale=0.35]{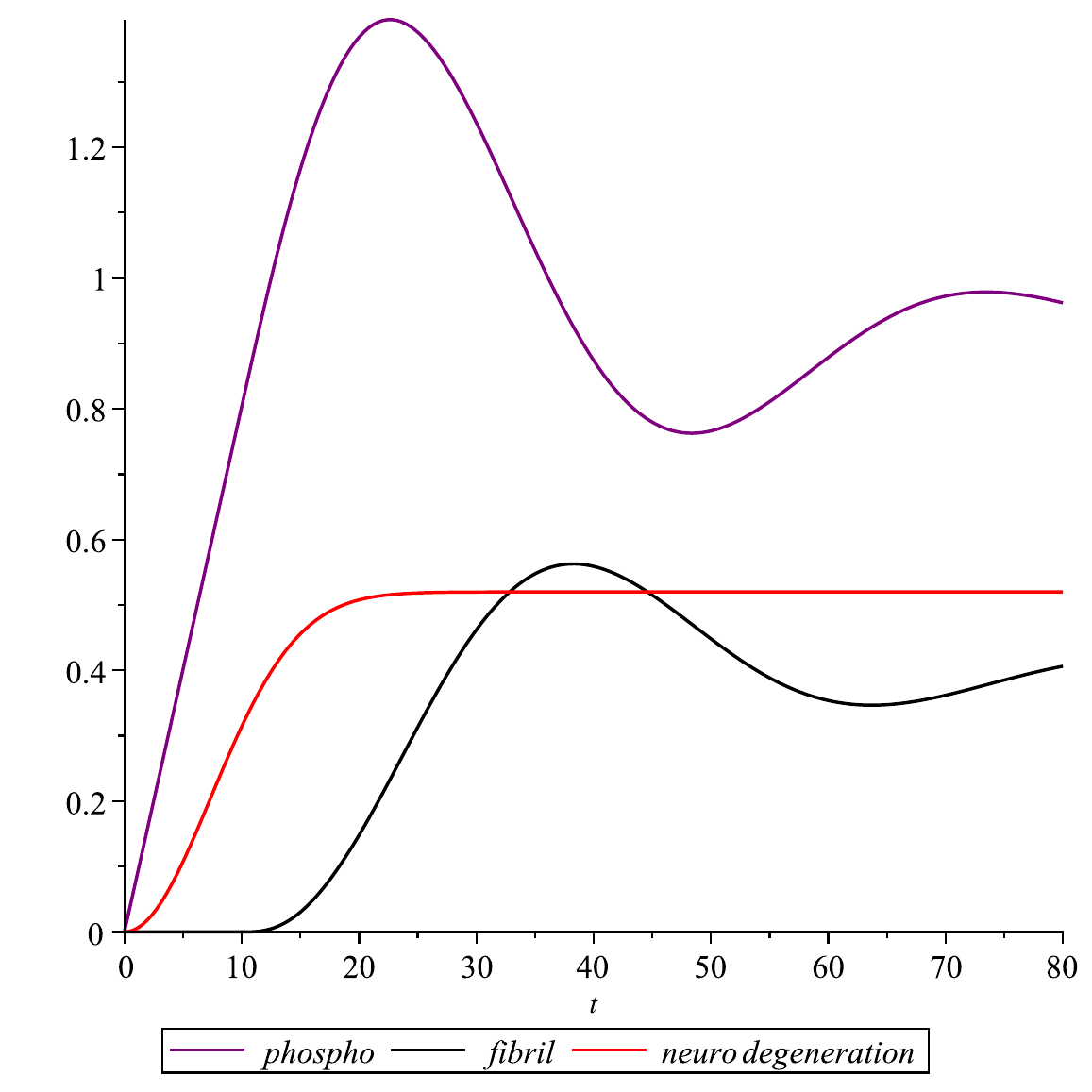}
\caption{Saturation of neuron degeneration  with no time lag between entangle and fibril production  }
 \end{figure}
 In case of almost no time lag, which means both fibril formation and entanglement is allowed to occur simultaneously, neuro degeneration function decays shortly shown by black line . 
 But, in presence of time lag [ Figure 2] ,tau filament increases significantly and oscillation of fibril  is  very significant.  In our analysis, degeneration function takes saturation value whereas 
 phosphorylated $ tau$ production in first phase increase sharply and then is marked by small oscillation with large time shift. We considered phosphorylated $\tau$ formation in secondary phase induced by amyloid$\beta$.

\end{document}